\documentclass[journal,12pt,onecolumn,draftclsnofoot,]{IEEEtran}
\usepackage{amsmath, graphics,amssymb,epsfig,subfigure}
\usepackage{cite}
\usepackage{url}
\usepackage{color,soul}
\usepackage{hhline}

\def\BibTeX{{\rm B\kern-.05em{\sc i\kern-.025em b}\kern-.08em
    T\kern-.1667em\lower.7ex\hbox{E}\kern-.125emX}}

\newtheorem{lemma}{Lemma}

\newfont{\bb}{msbm10 scaled 1000}

\newcommand{\av}{{\bf{a}}}

\newcommand{\cv}{{\bf{c}}}

\newcommand{\qv}{{\bf{q}}}

\newcommand{\uv}{{\bf{u}}}
\newcommand{\vv}{{\bf{v}}}
\newcommand{\wv}{{\bf{w}}}
\newcommand{\xv}{{\bf{x}}}

\newcommand{\be}{\begin{equation}}
\newcommand{\ee}{\end{equation}}
\newcommand{\bea}{\begin{eqnarray}}
\newcommand{\eea}{\end{eqnarray}}
\newcommand{\bitem}{\begin{itemize}}
\newcommand{\eitem}{\end{itemize}}

\makeatletter
\def\hlinewd#1{%
\noalign{\ifnum0=`}\fi\hrule \@height #1 \futurelet \reserved@a\@xhline}
\newcommand{\hthickline}{\hlinewd{.8pt}}

\makeatother

\makeatletter

\newcommand*{\rom}[1]{\expandafter\@slowromancap\romannumeral #1@}
\makeatother

\begin{document}

\title{UAV-Aided Wireless Communication Designs \\With Propulsion Energy Limitations}
\vspace{-30mm}

\author{\IEEEauthorblockN{Subin Eom, Hoon Lee, Junhee Park and Inkyu Lee, \textit{Fellow}, \textit{IEEE}} \\
\IEEEauthorblockA{School of Electrical Eng., Korea University, Seoul, Korea\\
    Email: \{esb777, ihun1, pjh0585, inkyu\}@korea.ac.kr}
 } \maketitle

\begin{abstract}
This paper studies unmanned aerial vehicle (UAV) aided wireless communication systems where a UAV supports uplink communications of multiple ground nodes (GNs) while flying over the area of the interest. In this system, the propulsion energy consumption at the UAV is taken into account so that the UAV's velocity and acceleration should not exceed a certain threshold. We formulate the minimum average rate maximization problem and the energy efficiency (EE) maximization problem by jointly optimizing the trajectory, velocity, and acceleration of the UAV and the uplink transmit power at the GNs. As these problems are non-convex in general, we employ the successive convex approximation (SCA) techniques. To this end, proper convex approximations for the non-convex constraints are derived, and iterative algorithms are proposed which converge to a local optimal point. Numerical results demonstrate that the proposed algorithms outperform baseline schemes for both problems. Especially for the EE maximization problem, the proposed algorithm exhibits about 109 $\%$ gain over the baseline scheme.

\end{abstract}
%
\section{Introduction} \label{sec:introduction}
Recently, unmanned aerial vehicles (UAVs) have received great attentions as a new communication entity in wireless networks \cite{YZeng:16}. Compared to conventional terrestrial communications where users are served by ground base stations (BSs) fixed at given position \cite{WLee:12}, UAV-aided systems could be dispatched to the field with various purposes such as disaster situations and military uses. Moreover, located high above users, UAVs are likely to have line-of-sight (LoS) communication links for air-to-ground channels.

Utilizing these advantages, UAVs have been considered to diverse wireless communication systems. The authors in \cite{YZeng:16_1} and \cite{QWang:17} studied a mobile relaying system where a UAV helps the communication of ground nodes (GNs) without direct communication links. In this UAV-aided relaying system, compared to conventional static relay schemes \cite{CSong:17},\cite{HKong:14}, the UAV can move closer to source and destination nodes in order to obtain good channel conditions, and thus the system throughput can be significantly improved. In \cite{YZeng:16_1}, the throughput of mobile relaying channels was maximized by optimizing the transmit power at the source and the relay node as well as the trajectory of the mobile relay. For the fixed relay trajectory, the work \cite{QWang:17} addressed the secrecy rate maximization problem for the UAV-based relaying system with an external eavesdropper.

In addition, UAVs have been adopted to assist conventional terrestrial communication infrastructures \cite{AMerwaday:15,JLyu:17_1,SJeong:17}. For the disaster situation, UAVs were employed in \cite{AMerwaday:15} to recover malfunctioned ground infrastructure. The work in  \cite{JLyu:17_1} examined a system where the UAV serves cell-edge users by jointly optimizing UAV's trajectory, bandwidth allocation, and user partitioning. Also, the flying computing cloudlets with UAVs were introduced to provide the offloading opportunities to multiple users \cite{SJeong:17}.

Moreover, the UAVs could play the role of mobile BSs in wireless networks \cite{AAl:14,JLyu:16,QWu:17}.
The authors in \cite{AAl:14} derived mathematical expressions for the optimum altitude of the UAVs that maximizes the coverage of the cellular network. Also, the trajectory optimization methods for mobile BSs were presented in \cite{JLyu:16} and \cite{QWu:17}. Assuming that the GNs are located in a line, the minimum throughput performance was maximized in \cite{JLyu:16} by optimizing the position of a UAV on a straight line. This result was extended in \cite{QWu:17} to a general scenario where multiple UAVs fly three-dimensional space to communicate with GNs. The joint optimization algorithms for the UAV trajectory, transmit power, and time allocation were provided in \cite{QWu:17} to maximize the minimum throughput performance. However, these works did not consider the propulsion energy consumption of the UAVs necessary for practical UAV designs under limited on-board energy situation \cite{filippone:06}.

By taking this issue into account, recent works \cite{DChoi:14_1,JZhang:17,YZeng:17} investigated energy efficiency (EE) of the UAV system.
Different from conventional systems which consider only communication-related energy consumption \cite{HKim:15,Jxu:13,SLee:16},
the EE of the UAV should addresses the propulsion energy at the UAV additionally. The authors in \cite{DChoi:14_1} maximized the EE by controlling the turning radius of a UAV for mobile relay systems. Also, by jointly optimizing the time allocation, speed, and trajectory, both the spectrum efficiency and the EE were maximized in \cite{JZhang:17}. In \cite{YZeng:17}, the propulsion energy consumption of the UAV was theoretically modeled, and the EE of the UAV was maximized for a single GN system.

This paper studies UAV-aided wireless communications where a UAV with limited propulsion energy receives the data of multiple GNs in the uplink. It is assumed that all GNs and the UAV operate in the same frequency band and there are no direct communication links among GNs. Under these setup, we formulate the minimum rate maximization problem and the EE maximization problem by jointly optimizing the UAV trajectory, the velocity, the acceleration, and the uplink transmit power at the GNs.
A similar approach for solving the minimum rate maximization was studied in \cite{QWu:17}, but the authors in \cite{QWu:17} did not involve the propulsion energy consumption at the UAV. For the EE maximization problem, our work can be regarded as a generalization of the single GN system in \cite{YZeng:17} to the multi-GN scenario, and thus we need to deal with inter-node interference as well. Due to these issues, existing algorithms presented in \cite{QWu:17} and \cite{YZeng:17} cannot be directly applied to our problems.

To tackle our problem of interest, we introduce auxiliary variables which couple the trajectory variables and the uplink transmit power in order to jointly optimize these variables. As the equivalent problem is still non-convex, we employ the successive convex approximation (SCA) technique which successively solves approximated convex problems of the original non-convex one. In order to apply the SCA to our optimization problems, we present new convex surrogate functions for the non-convex constraints. Then, we propose efficient algorithms for the minimum rate maximization problem and the EE maximization problem which yield local optimal solutions. Simulation results confirm that the proposed algorithms provide a significant performance gain over baseline schemes.

The rest of this paper is organized as follows: Section \rom{2} explains the system model and the problem formulations for the UAV-aided communication systems. In Section \rom{3}, the minimum rate maximization and the EE maximization algorithms are proposed. We examine the circular trajectory case as baseline schemes in Section \rom{4}. Section \rom{5} presents the numerical results for the proposed algorithms and we conclude the paper in Section \rom{6}.

$Notations$: Throughout this paper, the bold lower-case and normal letters denote vectors and scalars, respectively. The space of $M$-dimensional real-valued vectors are represented by  $\mathbb{R}^{M \times 1}$. For a vector $\av$, $\|\av\|$ and $\av^{T}$ indicate norm and transpose, respectively. The gradient of a function $f$ is defined as $\nabla f$. For a time-dependent function $\xv(t)$, $\dot{\xv}(t)$ and $\ddot{\xv}(t)$ stand for the first-order and second-order derivatives with respect to time $t$, respectively.

\section{System Model And Problem Formulation} \label{sec:system_model and problem formulation}

\begin{figure}
\begin{center}
\includegraphics[width=5in]{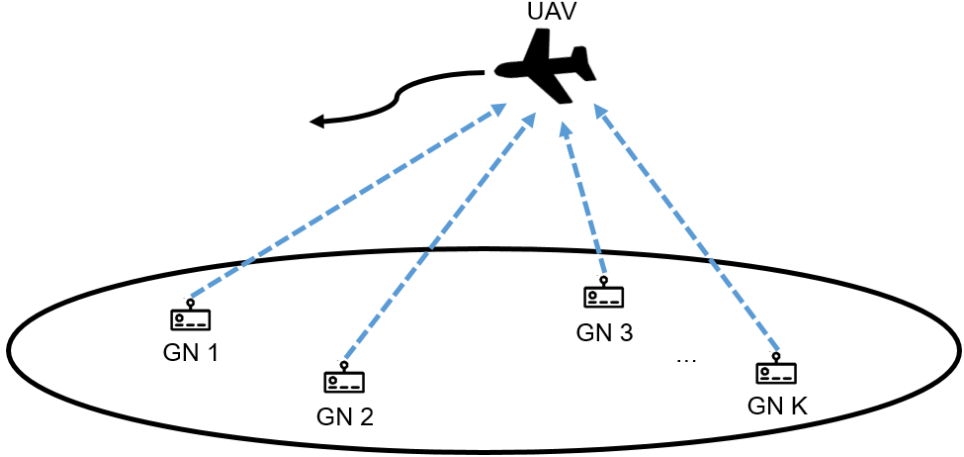}
\end{center}
\vspace{-5mm}
\caption{UAV-enabled wireless network}
\label{figure:system_model}
\end{figure}

As shown in Fig. 1, we consider UAV-aided wireless communications where a UAV receives uplink information transmitted from $K$ GNs. The UAV horizontally flies at a constant altitude $H$ with a time period $T$, while the GNs are located at fixed positions, which are perfectly known to the UAV in advance.
For the location of the GNs and the UAV, we employ a three-dimensional Cartesian coordinate system, and thus the horizontal coordinate of GN $k \ (k=1,...,K)$ is denoted by $\wv_{k}=[x_{k} \ y_{k}]^{T}$. Also, we define the time-varying horizontal coordinate of the UAV at time instant $t$ as $\qv(t)=[q_{x}(t) \ q_{y}(t)]^{T}, \ \text{for} \ 0\leq t \leq T$. Then, the instantaneous velocity $\vv(t)$ and the acceleration $\av(t)$ of the UAV are expressed by $\vv(t)\triangleq \dot{\qv}(t)$ and $\av(t)\triangleq \ddot{\qv}(t)$, respectively.

Continuous time expressions of variables make analysis and derivations in the UAV systems intractable. For ease of analysis, we discretize the time duration $T$ into $N$ time slots with the same time interval $\delta_{t}=\frac{T}{N}$ \cite{YZeng:16_1}. As a result, the trajectory of the UAV can be represented by $N$ vector sequences $\qv[n]\triangleq \qv(n\delta_{t})$, $\vv[n]\triangleq \vv(n\delta_{t})$, and $\av[n]\triangleq \av(n\delta_{t})$ for $n=0,1,...,N$. When the discretized time interval $\delta_{t}$ is chosen as a small number, the velocity and the acceleration can be approximated by using Taylor expansions as \cite{YZeng:17}
\bea
    &\vv[n]=\vv[n-1] + \av[n-1]\delta_{t},\ \text{for} \ \  n=1,...,N, \\
    &\qv[n]=\qv[n-1] + \vv[n-1]\delta_{t} + \frac{1}{2}\av[n-1]\delta^{2}_{t}, \ \text{for} \ \  n=1,...,N.
\eea
Also, assuming the periodical operation at the UAV, we have \cite{QWu:17}
\bea
    \qv[0]=\qv[N], \vv[0]=\vv[N], \av[0]=\av[N],
\eea
which implies that after one period $T$, the UAV returns to its starting location with the same velocity and acceleration.

In addition, the acceleration and the velocity of the practical UAV are subject to
\bea
    &\|\av[n]\|\leq a_{\text{max}},\ \text{for} \ \  n=0,1,...N, \\
    &V_{\text{min}}\leq\|\vv[n]\|\leq V_{\text{max}},\ \text{for} \ \  n=0,1,...,N,
\eea
where $a_{\text{max}}$ indicates the maximum UAV acceleration in m/sec$^{2}$ and $V_{\text{min}}$ and $V_{\text{max}}$ stand for the minimum and the maximum UAV speed constraints in m/sec, respectively. Notice that the minimum speed constraint $V_{\text{min}}$ is important for practical fixed-wing UAV designs which need to move forward to remain aloft and thus cannot hover over a fixed location \cite{YZeng:17}.

For the power consumption at the UAV, we take into account the propulsion power utilized for maintaining the UAV aloft and supporting its mobility. The propulsion power of the UAV $P_{\text{prop}}[n]$ at time slot $n$ is given by \cite{YZeng:17}
\bea
    P_{\text{prop}}[n]=c_{1}\|\vv[n]\|^{3}+\frac{c_{2}}{\|\vv[n]\|}\left(1+\frac{\|\av[n]\|^{2}}{g^{2}} \right),\ \text{for} \ \  n=0,1,...,N,
\eea
where $c_{1}$ and $c_{2}$ are the parameters related to the aircraft design and $g$ =  9.8 m/sec$^{2}$ equals the gravitational acceleration.
Thus, the average propulsion power and the total consumed propulsion energy over $N$ time slots are obtained by $\frac{1}{N}\sum_{n=1}^{N} P_{\text{prop}}[n]$ and $\delta_{t}\sum_{n=1}^{N} P_{\text{prop}}[n]$, respectively.
The power consumed by signal processing circuits such as analog-to-digital converters and channel decoders are ignored since they are practically much smaller than the propulsion power \cite{YZeng:17}.

Now, let us explain the channel model between the UAV and the GNs. We assume that the air-to-ground communication links are dominated by the LoS links. Moreover, the Doppler effect due to the UAV mobility is assumed to be well compensated. Then, the effective channel gain $h_{k}[n]$ from GN $k$ to the UAV at time slot $n$ follows the free-space path loss model as \cite{YZeng:16_1}
\bea
    h_{k}[n]=\frac{\gamma_{0}}{d_{k}^{2}[n]},
\eea
where $\gamma_{0} \triangleq \beta_{0}/\sigma^{2}$ represents the reference signal-to-noise ratio (SNR) at 1 m with $\beta_{0}$ and $\sigma^{2}$ being the channel power at 1 m and the white Gaussian noise power at the UAV, respectively, and the distance $d_{k}[n]$ is written by
\bea
    d_{k}[n]=\sqrt{\|\qv[n]-\wv_{k}\|^{2}+H^{2}}.
\eea

At time slot $n$, GN $k$ transmits its data signal to the UAV with power $0\leq p_{k}[n]\leq P_{\text{peak}}$, where $P_{\text{peak}}$ is the peak transmission power constraint at the GNs. Accordingly, the instantaneous achievable rate $R_{k}[n]$ can be expressed as
\bea
    R_{k}[n]=\log_{2}\left(1+\frac{p_{k}[n]h_{k}[n]}{1+\sum_{j=1,j\neq k}^{K}p_{j}[n]h_{j}[n] }\right),
\eea
where the term $\sum_{j=1,j\neq k}^{K}p_{j}[n]h_{j}[n] $ stands for interference from other GNs. Therefore, the achievable average rate of the GN $k$ and the total information bits transmitted from GN $k$ over $N$ time slots are denoted as $\frac{1}{N}\sum_{n=1}^{N}R_{k}[n]$ and $W\delta_{t}\sum_{n=1}^{N}R_{k}[n]$, respectively, where $W$ means the bandwidth.

In this paper, we jointly optimize the variables $\qv[n]$, $\vv[n]$, and $\av[n]$ and the uplink transmit power $p_{k}[n]$ at the GNs so that the minimum average rate among multiple GNs and the EE are maximized, respectively. First, the minimum rate maximization problem can be formulated as
\begin{subequations}
\begin{align}
    (P1) : &\max_{ \substack{    \{ \qv[n],\vv[n],\av[n] \} \\ \{p_{k}[n],\tau \}}  }\ \quad \quad \tau\\
    &\quad \quad s.t.\quad \quad \quad \quad \frac{1}{N}\sum_{n=1}^{N}R_{k}[n] \geq \tau, \forall k,\\
    & \quad \quad \quad \quad \quad \ \quad \quad 0\leq p_{k}[n]\leq P_{\text{peak}},  \forall k,n,\\
    & \quad \quad \quad \quad \quad \ \quad \quad \frac{1}{N}\sum_{n=1}^{N} P_{\text{prop}}[n] \leq P_{\text{lim}},     \\
    & \quad \quad \quad \quad \quad \ \quad \quad (1)-(5), \nonumber
\end{align}
\end{subequations}
where $P_{\text{lim}}$ in (10d) indicates the propulsion power constraint at the UAV.

Next, to support all of the individual GNs, the fairness based EE \cite{BDu:14,Yli:15,Yli:15_1} is more suitable than the network-wise EE \cite{Jxu:13,SLee:16}. Thus, we define the EE in the UAV-aided wireless communication systems as the ratio between the minimum information bits transmitted among the GNs and the total energy consumed at the UAV. Therefore, the EE maximization problem can be written by
\begin{subequations}
\begin{align}
    (P2) : &\max_{ \substack{    \{ \qv[n],\vv[n],\av[n] \} \\ \{p_{k}[n],\eta \}}  }\ \quad \ \ \frac{\eta}{\sum_{n=1}^{N} P_{\text{prop}}[n]}\\
    &\quad \quad \quad \quad \quad \quad \quad \ W\sum_{n=1}^{N}R_{k}[n] \geq \eta, \forall k,\\
    &\quad \quad s.t.\quad \quad \quad \ \ (1)-(5), (10c). \nonumber
\end{align}
\end{subequations}

In general, (P1) and (P2) are non-convex problems due to the constraints and the objective functions. Compared to \cite{QWu:17}, we additionally consider the propulsion power constraint (10d) in the minimum rate maximization problem (P1). Also, note that the EE maximization problem (P2) can be regarded as a generalization of \cite{YZeng:17} which investigated only a single GN scenario. From these respects, the works in \cite{QWu:17} and \cite{YZeng:17} can be regarded as special cases of our problems (P1) and (P2), respectively.
To solve the problems (P1) and (P2), we adopt the SCA framework \cite{BMarks:78} \cite{YSun:17} which iteratively solves approximated convex problems for the original non-convex problems.

\section{Proposed Algorithm } \label{sec:Proposed Algorithm For Unconstrained Trajectory}
In this section, we propose iterative algorithms for efficiently solving (P1) and (P2) by applying the SCA method. First, the minimum rate maximization problem (P1) is considered in Section \rom{3}-A, and then it is followed by the EE maximization problem (P2) in Section \rom{3}-B.

\subsection{Minimum Average Rate Maximization}
Applying the change of variables as
\bea
    G_{k}[n]\triangleq p_{k}[n]h_{k}[n]=\frac{p_{k}[n]\gamma_{0}}{ {\|\qv[n]-\wv_{k} \|}^{2} +H^{2} }, \forall k, n,
\eea
where $G_{k}[n]$ is a new optimization variable,
the constraint (10c) becomes $0\leq G_{k}[n]\leq G_{k,\text{max}}[n],  \forall k,n$, where $G_{k,\text{max}}[n]\triangleq P_{\text{peak}}h_{k}[n]=\frac{P_{\text{peak}}\gamma_{0}}{ {\|\qv[n]-\wv_{k} \|}^{2} +H^{2} }.$
Then, we can rewrite the achievable rate $R_{k}[n]$ in (9) as
\bea
    R_{k}[n]=\log_{2}\left(1+\sum_{m=1}^{K}G_{m}[n] \right) - \hat{R}_{k}[n],
\eea
where $\hat{R}_{k}[n] \triangleq \log_{2}\left(1+\sum_{j=1,j\neq k}^{K}G_{j}[n]\right).$

By introducing new auxiliary variables $\{V_{1}[n] \}$, (P1) can be recast to
\begin{subequations}
\begin{align}
    (P1.1) : &\max_{ \substack{    \{ \qv[n],\vv[n],\av[n] \} \\ \{G_{k}[n],V_{1}[n],\tau \}}  }\ \quad \quad \tau\\
    &\quad \quad s.t.\quad \quad \quad \quad \frac{1}{N}\sum_{n=1}^{N}\left(\log_{2}\left(1+\sum_{m=1}^{K}G_{m}[n] \right) - \hat{R}_{k}[n]\right) \geq \tau, \forall k,\\
    & \quad \quad \quad \quad \quad \quad \quad \ 0\leq G_{k}[n]\leq G_{k,\text{max}}[n],  \forall k,n,\\
    & \quad \quad \quad \quad \quad \quad \quad \ \frac{1}{N}\sum_{n=1}^{N} c_{1}\|\vv[n]\|^{3}+\frac{c_{2}}{V_{1}[n]}+\frac{c_{2}\|\av[n]\|^{2}}{g^{2}V_{1}[n]} \leq P_{\text{lim}},     \\
    & \quad \quad \quad \quad \quad \quad \quad \ V_{\text{min}} \leq V_{1}[n],  \forall n, \\
    & \quad \quad \quad \quad \quad \quad \quad \ V_{1}^{2}[n] \leq {\| \vv[n]\|}^2,  \forall n, \\
    & \quad \quad \quad \quad \quad \quad \quad \ \|\vv[n]\|\leq V_{\text{max}}, \forall n, \\
    & \quad \quad \quad \quad \quad \quad \quad \ (1)-(4). \nonumber
\end{align}
\end{subequations}

It can be shown that at the optimal point of (P1.1), the inequality constraint in (14f) holds with the equality, since otherwise we can enlarge the feasible region corresponding to (14d) by increasing $V_{1}[n]$. Therefore, we can conclude that (P1.1) is equivalent to (P1). Thanks to the new auxiliary variables $\{V_{1}[n]\}$, constraints (14d) and (14e) now become convex, while (14b), (14c), and (14f) are still non-convex in general.

To address these constraints, we employ the SCA methods. First, it can be checked that constraint (14b) is given by a difference of two concave functions. Hence, the convex surrogate function $\hat{R}_{k}^{\text{ub}}[n]$ for $\hat{R}_{k}[n]$ can be computed from a first order Taylor approximation as
\bea
    \hat{R}_{k}^{\text{ub}}[n] \triangleq  \hat{\Gamma}_{k}[n]\sum_{j=1,j\neq k}^{K}\left(   G_{j,l+1}[n] - G_{j,l}[n]     \right)       +  \log_{2}\left(1+\sum_{j=1,j\neq k}^{K}G_{j,l}[n]\right) \geq \hat{R}_{k}[n]  ,
\eea
where $G_{k,l}[n]$ indicates a solution of $G_{k}[n]$ attained at the $l$-th iteration of the SCA process and $\hat{\Gamma}_{k}[n] \triangleq \log_{2}e / (1+\sum_{j=1,j\neq k}^{K}G_{j,l}[n]).$
Next, to identify the surrogate functions of (14c) and (14f), we present the following lemmas.
\begin{lemma}
Denoting $\{\qv_{l}[n]\}$ as a solution for $\{\qv[n]\}$ calculated at the $l$-th iteration, the concave surrogate function $G_{k,\text{max}}^{\text{lb}}[n]$ for $G_{k,\text{max}}[n]$ can be expressed as
\bea
    G_{k,\text{max}}^{\text{lb}}[n]&\triangleq&P_{\text{peak}}\gamma_{0} \left(   -\frac{ {\|\qv_{l+1}[n]-\wv_{k} \|}^{2}  }{H^4} +   B_{k}[n]{ \left(   \qv_{l+1}[n]-\wv_{k}  \right)   }^{T}{ \left(    \qv_{l}[n]-\wv_{k}   \right)  }  +  C_{k}[n]                     \right) \nonumber \\
    &\leq&G_{k,\text{max}}[n],
\eea
where the constants $B_{k}[n]$ and $C_{k}[n]$ are respectively given as
\bea
    B_{k}[n] &\triangleq& 2\left( \frac{1}{H^4} - \frac{1}{ {\left( {\|\qv_{l}[n]-\wv_{k} \|}^{2} + H^2 \right)}^2 } \right),  \nonumber \\
    C_{k}[n] &\triangleq& \frac{1}{ { \| \qv_{l}[n]-\wv_{k} \| }^{2} + H^2 } + \frac{ 2 { {\|\qv_{l}[n]-\wv_{k} \|}^{2} } }{ {\left( {\|\qv_{l}[n]-\wv_{k} \|}^{2} + H^2 \right)}^2 } - \frac{{\|\qv_{l}[n]-\wv_{k} \|}^{2}}{H^4}. \nonumber
\eea
\end{lemma}
\begin{IEEEproof}
Please refer to Appendix A.
\end{IEEEproof}
\begin{lemma}
From a solution $\{\vv_{l}[n]\}$ obtained at the $l$-th iteration, the concave surrogate function of $\| \vv_{l+1}[n] \|^{2}$ can be computed as
\bea
     -\|\vv_{l+1}[n] \|^{2} + 2\vv_{l}^{T}\left(2\vv_{l+1}[n] - \vv_{l}[n] \right) \leq \| \vv_{l+1}[n] \|^{2}.
\eea
\end{lemma}
\begin{IEEEproof}
Applying a similar process in Appendix A, we can conclude that the function in (17) satisfies the conditions for a concave surrogate function \cite{BMarks:78}.
\end{IEEEproof}

With the aid of Lemmas 1 and 2, at the $(l+1)$-th iteration, the non-convex constraints in (14c) and (14f) can be approximated as
\bea
    &0\leq G_{k}[n]\leq G_{k,\text{max}}^{\text{lb}}[n], \\
    &V_{1}^{2}[n] \leq -\|\vv_{l+1}[n] \|^{2} + 2\vv_{l}^{T}\left(2\vv_{l+1}[n] - \vv_{l}[n] \right).
\eea
As a result, with given solutions $\{\qv_{l}[n]$, $\vv_{l}[n]$, $G_{k,l}[n]\}$ at the $l$-th iteration, we solve the following problem at the $(l+1)$-th iteration of the SCA procedure
\begin{subequations}
\begin{align}
    (P1.2) : &\max_{ \substack{    \{ \qv_{l+1}[n],\vv_{l+1}[n],\av[n] \} \\ \{G_{k,l+1}[n],V_{1}[n],\tau^{\text{lb}} \}}  }\ \quad \quad \tau^{\text{lb}}\\
    & \quad \quad \quad s.t.\quad \quad \quad \quad \quad \frac{1}{N}\sum_{n=1}^{N}\left(\log_{2}\left(1+\sum_{m=1}^{K}G_{m,l+1}[n] \right) - \hat{R}_{k}^{\text{ub}}[n]\right) \geq \tau^{\text{lb}}, \forall k, \\
    & \quad \quad \quad \quad \quad \quad \quad \quad \quad \ (1)-(4),(14d),(14e),(14g),(18),(19), \nonumber
\end{align}
\end{subequations}
where $\tau^{\text{lb}}$ denotes the lower bound of $\tau$ in the original problem (P1).
Since (P1.2) is a convex problem, it can be optimally solved via existing convex optimization solvers, e.g. CVX \cite{CVX:17}. Based on these results, we summarize the proposed iterative procedure in Algorithm 1.
\renewcommand{\arraystretch}{.79}
\begin{center}
\begin{tabular}{l}
\hthickline
\textbf{Algorithm 1}: Proposed algorithm for (P1) \\
\hthickline
Initialize $\{\qv_{0}[n], \vv_{0}[n], G_{k,0}[n]\},  \forall k,n$ and let $l=0$.\\
\textbf{Repeat}\\
   \makebox[20pt]{   }Compute $\{\qv_{l+1}[n], \vv_{l+1}[n], G_{k,l+1}[n]\}$ for (P1.2) with given $\{\qv_{l}[n], \vv_{l}[n], G_{k,l}[n]\}$. \\
   \makebox[20pt]{   }Update $l \leftarrow l+1$.\\
\textbf{Until} Convergence.\\
Obtain $p_{k}[n]=\frac{G_{k,l+1}[n]   }{h_{k,l+1}[n]}$.\\
\hthickline
\end{tabular}
\end{center}

For the convergence analysis of Algorithm 1, let us define the objective values of (P1) and (P1.2) at the $l$-th iteration as $\tau_{l}$ and $\tau^{\text{lb}}_{l}$, respectively. Then we can express the relationship
\bea
    \tau_{l} \ = \ \tau^{\text{lb}}_{l} \ \leq \ \tau^{\text{lb}}_{l+1} \leq \  \tau_{l+1},
\eea
where the first equation holds because the surrogate functions in (15), (16), and (17) are tight at the given local points, the second inequality is derived from the non-decreasing property of the optimal solution of (P1.2), and the third inequality follows from the fact that the approximation problem (P1.2) is a lower bound of the original problem (P1).

From (21), we can conclude that the objective value $\tau$ in (P1) is non-decreasing for every iterations of Algorithm 1. Since the objective value $\tau$ in (P1) has a finite upper bound value and at given local points, the surrogate functions in (15), (16), and (17) obtain the same gradients as their original functions, it can be verified that Algorithm 1 is guaranteed to converge to at least a local optimal solution for (P1) \cite{BMarks:78,YSun:17}.
\subsection{Energy Efficiency Maximization}
In this subsection, we consider the EE maximization problem (P2). First, by applying (12)-(13), and introducing an auxiliary variable $\{V_{1}[n]\}$, (P2) can be transformed as
\begin{subequations}
\begin{align}
    (P2.1) : &\max_{\substack{    \{ \qv[n],\vv[n],\av[n] \} \\ \{G_{k}[n],V_{1}[n],\eta \}}  }\ \quad \quad \frac{\eta}{\sum_{n=1}^{N} c_{1}\|\vv[n]\|^{3}+\frac{c_{2}}{V_{1}[n]}+\frac{c_{2}\|\av[n]\|^{2}}{g^{2}V_{1}[n]} } \\
    &\quad \quad s.t.\quad \quad \quad \quad \ W\sum_{n=1}^{N}\left(\log_{2}\left(1+\sum_{m=1}^{K}G_{m}[n] \right) - \hat{R}_{k}[n]\right) \geq \eta, \forall k,\\
    & \quad \quad \quad \quad \quad \quad \quad \ \ (1)-(4),(14c),(14e)-(14g). \nonumber
\end{align}
\end{subequations}
Similar to (P1.1), we can see that (P2.1) is equivalent to (P2), but (P2.1) is still non-convex due to the constraints in (14c), (14f), and (22b).

To tackle this issue, we can employ the similar SCA process presented in Section \rom{3}-A. By adopting (15) and Lemmas 1 and 2, a convex approximation of (P2.1) at the $(l+1)$-th iteration is given by
\begin{subequations}
\begin{align}
    (P2.2) : &\max_{ \substack{    \{ \qv_{l+1}[n],\vv_{l+1}[n],\av[n] \} \\ \{G_{k,l+1}[n],V_{1}[n],\eta^{\text{lb}} \}}  }\ \quad \quad \frac{\eta^{\text{lb}}}{\sum_{n=1}^{N} c_{1}\|\vv[n]\|^{3}+\frac{c_{2}}{V_{1}[n]}+\frac{c_{2}\|\av[n]\|^{2}}{g^{2}V_{1}[n]} } \\
    & \quad \quad \quad s.t.\quad \quad \quad \quad \quad \ W\sum_{n=1}^{N}\left(\log_{2}\left(1+\sum_{m=1}^{K}G_{m,l+1}[n] \right) - \hat{R}_{k}^{\text{ub}}[n]\right) \geq \eta^{\text{lb}}, \forall k \\
    & \quad \quad \quad \quad \quad \quad \quad \quad \quad \ \ (1)-(4),(14e),(14g),(18),(19), \nonumber
\end{align}
\end{subequations}
where $\eta^{\text{lb}}$ denotes the lower bound of $\eta$ in the original problem (P2).

It can be shown that (P2.2) is a concave-convex fractional problem, which can be optimally solved via the Dinkelbach's method \cite{WDinkelbach:67},\cite{AZappone:15}.
Then, denoting $\mu = \sum_{n=1}^{N} c_{1}\|\vv[n]\|^{3}+\frac{c_{2}}{V_{1}[n]}+\frac{c_{2}\|\av[n]\|^{2}}{g^{2}V_{1}[n]} $ with a given constant $\lambda_{m}$, (P2.2) can be converted to (P2.3) as
\begin{subequations}
\begin{align}
    (P2.3) : &\max_{ \substack{    \{ \qv_{l+1}[n],\vv_{l+1}[n],\av[n] \} \\ \{G_{k,l+1}[n],V_{1}[n],\eta^{\text{lb}} \}}  }\ \quad \quad \eta^{\text{lb}} - \lambda_{m}\mu \\
    & \quad \quad \quad s.t.\quad \quad \quad \quad \quad  (1)-(4),(14e),(14g),(18),(19),(23b). \nonumber
\end{align}
\end{subequations}
Based on (P2.3), we summarize the proposed iterative procedure in Algorithm 2. The convergence and the local optimality of Algorithm 2 can be verified similar to Algorithm 1, and thus the details are omitted for brevity.
\renewcommand{\arraystretch}{.79}
\begin{center}
\begin{tabular}{l}
\hthickline
\textbf{Algorithm 2}: Proposed algorithm for (P2) \\
\hthickline
Initialize $\{\qv_{0}[n], \vv_{0}[n], G_{k,0}[n]\},  \forall k,n$ and let $\lambda_{0}=0$, $m=0$, and $l=0$.\\
\textbf{Repeat}\\
   \makebox[20pt]{   }\textbf{Repeat}\\
\makebox[20pt]{   }   \makebox[20pt]{   }Compute $\{\qv_{l+1}[n], \vv_{l+1}[n], G_{k,l+1}[n]\}$ for (P2.3) with given  \\
\makebox[20pt]{   }   \makebox[20pt]{   }$\{\qv_{l}[n], \vv_{l}[n], G_{k,l}[n]\},  \forall k,n $ and $\lambda_{m}$. \\
\makebox[20pt]{   }   \makebox[20pt]{   }Update $l \leftarrow l+1$.\\
   \makebox[20pt]{   }\textbf{Until} Convergence.\\
   \makebox[20pt]{   }Let $F(\lambda_{m}) = \eta^{lb}- \lambda_{m}\mu $ and $\lambda_{m+1} = \eta^{lb} / \mu $.\\
   \makebox[20pt]{   }Update $m \leftarrow m+1$.\\
   \makebox[20pt]{   }Let $\{\qv_{0}[n], \vv_{0}[n], G_{k,0}[n]\}$ = $\{\qv_{l+1}[n], \vv_{l+1}[n], G_{k,l+1}[n]\},  \forall k,n$ and $l = 0$. \\
\textbf{Until} Convergence.\\
Obtain $p_{k}[n]=\frac{G_{k,l+1}[n]   }{h_{k,l+1}[n]},  \forall k,n$.\\
\hthickline
\end{tabular}
\end{center}
It is worthwhile to note that we need to initialize the trajectory variables $\{\qv[n],\vv[n] \}$ for (P1) and (P2). However, it is not trivial to find such variables satisfying the UAV movement constraints (1)-(5) and the propulsion power constraint (10d). This will be clearly explained in Section \rom{4}-C.

\section{Circular trajectory system} \label{sec:Proposed Algorithm For Circular Trajectory}
Now, we examine the circular trajectory system which will be used as a baseline scheme. First, we choose the center of the circular trajectory $\cv=[x_{0} \ y_{0}]^{T}$ as the geometrical mean of the GNs $\cv = \frac{\sum_{k=1}^{K}\wv_{k}}{K}$. Denoting $r$ as the radius of the trajectory and $\theta[n]$ as the angle of the circle along which the UAV flies at time slot $n$, the horizontal coordinate of the UAV $\qv[n]$ can be obtained by $\qv[n] = [r\cos\theta[n] + x_{0} \ \ r\sin\theta[n] + y_{0}  ]^{T}$. Also, the location of GN $k$ $\wv_{k}$ can be represented as $\wv_{k}   = [\zeta_{k}\cos\varphi_{k} + x_{0} \ \ \zeta_{k}\sin\varphi_{k}+y_{0} ]^{T} $, where $\zeta_{k}$ and $\varphi_{k}$ equal the distance and the angle between the geometric center $\cv$ and GN $k$, respectively.
Thus, the distance $d_{k}[n]$ between the UAV and GN $k$ in (8) can be expressed as $d_{k}[n]=\sqrt{ r^{2}+\zeta_{k}^{2} + H^{2} -2r\zeta_{k}\cos\left( \theta[n] - \varphi_{k} \right)    }.$

By adopting the angular velocity $\omega[n]$ and the angular acceleration $\alpha[n]$, equations in (1)-(6) can be rewritten as
\bea
    &\omega[n]=\omega[n-1] + \alpha[n-1]\delta_{t},\ \text{for} \ \  n=1,...,N, \\
    &\theta[n]=\theta[n-1] + \omega[n-1]\delta_{t} + \frac{1}{2}\alpha[n-1]\delta^{2}_{t},\ \text{for} \ \  n=1,...,N, \\
    &\theta[N]=\theta[0] + 2\pi, \omega[0]=\omega[N], \alpha[0]=\alpha[N], \\
    &\|\av[n]\|^{2}={\|\av_{\parallel}[n]\|}^{2}+{\|\av_{\perp}[n]\|}^{2} = r^{2}\alpha^{2}[n] + r^{2}\omega^{4}[n] \leq a_{\text{max}}^{2},\ \text{for} \ \   n=0,1,...N, \\
    &\omega_{\text{min}} \leq \omega[n] \leq \omega_{\text{max}},\ \text{for} \ \  n=0,1,...,N, \\
    &P_{\text{prop}}[n] = c_{1}r^{3}\omega^{3}[n]+\frac{c_{2}}{r\omega[n]} + \frac{c_{2}r\omega^{3}[n]}{g^2}+\frac{c_{2}r\alpha^{2}[n]}{g^{2}\omega[n]},\ \text{for} \ \  n=0,1,...,N,
\eea
where $\av_{\parallel}[n]$ and $\av_{\perp}[n]$ are the tangential and centripetal accelerations, respectively, and $\omega_{\text{min}} \triangleq V_{\text{min}}/r$ and $\omega_{\text{max}} \triangleq V_{\text{max}}/r$ indicate the minimum and maximum angular velocity, respectively.

Similar to Section \rom{3}, we address the minimum average rate maximization problem and the EE maximization problem for the circular trajectory, which are respectively formulated as
\begin{subequations}
\begin{align}
    (P3) : &\max_{ \substack{    \{ \theta[n],\omega[n],\alpha[n] \} \\ \{ r,p_{k}[n],\tau \}}  }\ \quad \quad \tau\\
    &\quad \quad s.t.\quad \quad \quad \quad  r_{\text{min}} \leq r \leq r_{\text{max}}, \\
    & \quad \quad \quad \quad \quad \quad \quad \ (10b)-(10d),(25)-(29), \nonumber
\end{align}
\end{subequations}
\begin{subequations}
\begin{align}
    (P4) : &\max_{ \substack{    \{ \theta[n],\omega[n],\alpha[n] \} \\ \{ r,p_{k}[n],\eta \}}  }\ \quad \ \ \frac{\eta}{\sum_{n=1}^{N} P_{\text{prop}}[n]}\\
    &\quad \quad s.t.\quad \quad \quad \ \ (10c),(11b),(25)-(29),(31b), \nonumber
\end{align}
\end{subequations}
where $r_{\text{min}} \triangleq \frac{V_{\text{min}}T}{2\pi}$ and $r_{\text{max}} \triangleq \min \left( \frac{V_{\text{max}}T}{2\pi},\frac{a_{\text{max}}}{\max(\sqrt{\omega^{4}[n] + \alpha^{2}[n])} }  \right)$ denote the minimum and maximum radius of the circular trajectory, respectively.
It is emphasized that (P3) and (P4) are difficult to solve because of the non-convex constraints and objective functions. To deal with the problems (P3) and (P4), similar SCA frameworks in Section \rom{3} are applied.

\subsection{Minimum Average Rate Maximization and EE maximization}
For the minimum average rate maximization problem (P3), we first find $\{r, p_{k}[n]\}$ with given $\{\theta[n], \omega[n], \alpha[n]\}$ and then updates $\{\theta[n], \omega[n], \alpha[n], p_{k}[n]\}$ for a fixed $r$.
For given $\{\theta[n], \omega[n], \alpha[n]\}$, we adopt the change of variable $S_{k}[n]$ and $S_{k,\text{max}}[n]$ as
\bea
    S_{k}[n] \triangleq p_{k}[n]h_{k}[n] = \frac{p_{k}[n]\gamma_{0}}{{\left( r-\zeta_{k}\cos\left( \theta[n] - \theta_{k} \right) \right)}^2 + \zeta_{k}^{2}\sin^{2}\left( \theta[n] - \theta_{k} \right) + H^2},
\eea
\bea
    S_{k,\text{max}}[n] \triangleq P_{\text{peak}}h_{k}[n] = \frac{P_{\text{peak}}\gamma_{0}}{{\left( r-\zeta_{k}\cos\left( \theta[n] - \theta_{k} \right) \right)}^2 + \zeta_{k}^{2}\sin^{2}\left( \theta[n] - \theta_{k} \right) + H^2}.
\eea


Similar to the method in Section \rom{3}-A, we employ the SCA to $S_{k,\text{max}}[n]$. Based on Lemma 1, the concave surrogate function $S_{k,\text{max}}^{\text{lb1}}[n]$ of $S_{k,\text{max}}[n]$ with a solution $r_{l}$ at the $l$-th iteration can be chosen as
\bea
    S_{k,\text{max}}^{\text{lb1}}[n]&\triangleq&P_{\text{peak}}\gamma_{0} \left(   -\frac{ {\left(  r_{l+1} - \check{b}_{k}[n] \right)}^{2}  }{\check{A}_{k}^{2}[n]} +   \check{B}_{k}[n]{ \left(   r_{l+1}-\check{b}_{k}[n]  \right)   }{ \left( r_{l}-\check{b}_{k}[n]    \right)  }  +  \check{C}_{k}[n]                     \right) \nonumber \\
    &\leq&S_{k,\text{max}}[n],  \forall n,
\eea
where the constants $\check{b}_{k}[n]$, $\check{A}_{k}[n]$, $\check{B}_{k}[n]$, and $\check{C}_{k}[n]$ are respectively given by
\bea
    &\check{b}_{k}[n] \triangleq \zeta_{k}\cos\left( \theta[n] - \theta_{k}\right), \nonumber\\
    &\check{A}_{k}[n] \triangleq \zeta_{k}^{2}\sin^{2}\left( \theta[n]-\theta_{k} \right) + H^2,\nonumber
\eea
\bea
    \check{B}_{k}[n] \triangleq 2\left(  \frac{1}{\check{A}_{k}^{2}[n]}  - \frac{1}{     {\left(    {\left(    r_{l}-\check{b}_{k}[n]     \right)}^2         +\check{A}_{k}[n]  \right)}^2         }   \right), \nonumber
\eea
\bea
    \check{C}_{k}[n] \triangleq \frac{1}{    {\left(    r_{l}-\check{b}_{k}[n]     \right)}^2         +\check{A}_{k}[n]  } + \frac{2{\left(    r_{l}-\check{b}_{k}[n]     \right)}^2}{    {\left(    {\left(    r_{l}-\check{b}_{k}[n]     \right)}^2         +\check{A}_{k}[n]  \right)}^2     } - \frac{{\left(    r_{l}-\check{b}_{k}[n]     \right)}^2}{\check{A}_{k}^{2}[n]}. \nonumber
\eea

By applying (15), (P3) for fixed $\{\theta[n], \omega[n], \alpha[n]\}$ can be reformulated as an approximated convex problem at the $(l+1)$-th iteration of the SCA
\begin{subequations}
\begin{align}
    (P3.1) : &\max_{ \{ r_{l+1},S_{k,l+1}[n],\tau^{\text{lb1}} \} } \quad  \ \tau^{\text{lb1}}\\
    &\quad \quad \quad  s.t. \quad \quad \quad \ \ \frac{1}{N}\sum_{n=1}^{N}\left(\log_{2}\left(1+\sum_{m=1}^{K}S_{m,l+1}[n] \right) - \breve{R}_{k}^{\text{ub}}[n]\right) \geq \tau^{\text{lb1}}, \forall k,\\
    &\quad  \quad \quad \quad \quad \quad \quad \ \ \ 0\leq S_{k}[n]\leq S_{k,\text{max}}^{\text{lb1}}[n],  \forall k,n,\\
    &\quad  \quad \quad \quad \quad \quad \quad \ \ \ (10d), (31b), \nonumber
\end{align}
\end{subequations}
where $\breve{R}_{k}^{\text{ub}}[n] \triangleq  \breve{\Gamma}_{k}[n] \left( \sum_{j=1,j\neq k}^{K}\left(   S_{j,l+1}[n] - S_{j,l}[n]     \right)       \right) +  \log_{2}\left(1+\sum_{j=1,j\neq k}^{K}S_{j,l}[n]\right)$ and $\breve{\Gamma}_{k}[n] \triangleq \frac{\log_{2}e}{1+\sum_{j=1,j\neq k}^{K}S_{j,l}[n]}.$ (P3.1) can be successively solved by the CVX until convergence.

Next, we present a solution for (P3) with a given $r$.
To obtain the concave surrogate function of $S_{k,\text{max}}[n]$, we introduce the following lemma which identifies the surrogate function of the cosine function.

\begin{lemma}
For any given $\phi_{l}$, the concave surrogate function of $\cos \phi$ can be computed as
\bea
     \frac{-{\left(  \phi-\phi_{l}+\sin \phi_{l}     \right)}^2}{2} + \cos \phi_{l} + \frac{\sin^{2}\phi_{l}}{2} \leq \cos \phi .
\eea
\end{lemma}
\begin{IEEEproof}
With a similar process in Appendix A, we can conclude that the function in (37) satisfies the conditions for a concave surrogate function \cite{BMarks:78}.
\end{IEEEproof}

By inspecting Lemmas 1 and 3, the concave surrogate function $S_{k,\text{max}}^{\text{lb2}}[n]$ for $S_{k,\text{max}}[n]$ can be identified as
\bea
    S_{k,\text{max}}^{\text{lb2}}[n] \triangleq P_{\text{peak}}\gamma_{0} \left(   -\frac{ r\zeta_{k}{\left( \theta_{l+1}[n] - \hat{b}_{k}[n]\right)}^2   }{ \hat{A}_{k}^{2}[n]  } +\hat{B}_{k}[n]\sin(\theta_{l}[n]-\theta_{k})\left( \theta_{l+1}[n]-\hat{b}_{k}[n]\right) + \hat{C}_{k}[n]       \right) \nonumber
\eea
\bea
    \leq \frac{P_{\text{peak}}\gamma_{0}}{r\zeta_{k}{\left( \theta_{l+1}[n] - \hat{b}_{k}[n] \right)}^2 + \hat{A}_{k}[n]} \leq S_{k,\text{max}}[n], \quad \quad \quad \quad \quad \quad \quad \quad \quad \quad
\eea
where $\hat{b}_{k}[n]$, $\hat{A}_{k}[n]$, $\hat{B}_{k}[n]$, and $\hat{C}_{k}[n]$ are given by
\bea
    &\hat{b}_{k}[n] \triangleq \theta_{l}[n] - \sin\left( \theta_{l}[n] - \theta_{k}\right), \nonumber \\
    &\hat{A}_{k}[n] \triangleq r^{2} + \zeta_{k}^{2} + H^2 -r\zeta_{k}\left(  2\cos\left(  \theta_{l}[n] - \theta_{k} \right)   + \sin^{2}\left(  \theta_{l}[n] - \theta_{k}  \right) \right), \nonumber
\eea
\bea
    \hat{B}_{k}[n] \triangleq 2r\zeta_{k}\left(  \frac{1}{\hat{A}_{k}^{2}[n]}  - \frac{1}{     {\left(    r\zeta_{k}\sin^{2}(\theta_{l}[n]-\theta_{k})         +\hat{A}_{k}[n]  \right)}^2         }   \right), \quad \quad \quad \quad \quad \quad \quad \quad \quad \nonumber\\
    \hat{C}_{k}[n] \triangleq \frac{1}{   r\zeta_{k}\sin^{2}(\theta_{l}[n]-\theta_{k})         +\hat{A}_{k}[n] } + \frac{2r\zeta_{k}\sin^{2}(\theta_{l}[n]-\theta_{k})}{    {\left(    r\zeta_{k}\sin^{2}(\theta_{l}[n]-\theta_{k})         +\hat{A}_{k}[n]  \right)}^2     } - \frac{r\zeta_{k}\sin^{2}(\theta_{l}[n]-\theta_{k})}{\hat{A}_{k}^{2}[n]}. \quad \nonumber
\eea

By utilizing (15) and (38), at the $(l+1)$-th iteration of the SCA algorithm with a given $r$, (P3) can be approximated to the following convex problem.

\begin{subequations}
\begin{align}
    (P3.2) : &\max_{ \substack{    \{ \theta_{l+1}[n],\omega[n],\alpha[n] \} \\ \{ S_{k,l+1}[n],\tau^{\text{lb2}} \}}  }\ \quad \tau^{\text{lb2}}\\
    &\quad \quad \ s.t.\quad \quad \quad \ \ \frac{1}{N}\sum_{n=1}^{N}\left(\log_{2}\left(1+\sum_{m=1}^{K}S_{m,l+1}[n] \right) - \breve{R}_{k}^{\text{ub}}[n]\right) \geq \tau^{\text{lb2}}, \forall k,\\
    & \quad \quad \quad \quad \quad \quad \quad \ 0\leq S_{k}[n]\leq S_{k,\text{max}}^{\text{lb2}}[n],  \forall k,n,\\
    & \quad \quad \quad \quad \quad \quad \quad \  (10d),(25)-(29). \nonumber
\end{align}
\end{subequations}
We then successively solve (P3.2) by the CVX until convergence. Similar to Algorithm 1, a solution of problem (P3) is obtained by alternately solving (P3.1) and (P3.2) until the objective value converges.

For the EE maximization problem (P4) in the circular trajectory case, we can apply similar methods in Section \rom{3}-B. Based on (P3.1) and (P3.2), given $\{\theta[n], \omega[n], \alpha[n]\}$ and $r$, (P4) can be transformed into two concave-convex fractional problems. By using Algorithm 2, we can alternately solve these problems until convergence.

\subsection{Trajectory Initialization}
To initialize the proposed algorithms, we employ a simple circular path concept in \cite{QWu:17}. First, the initial angular velocity $\omega_{0}$ is set to $\omega_{0} = \frac{2\pi}{T}$, which implies $\theta_{0}[n]=2\pi\frac{n}{N}$, $\forall n$. Next, the initial radius $r_{0}$  is chosen to fulfill the constraints in (4), (5), and (10d), which can be expressed as
\bea
    &\frac{V_{\text{min}}T}{2\pi}\leq r_{0} \leq \min \left( \frac{V_{\text{max}}T}{2\pi},\frac{a_{\text{max}}}{\omega_{0}^{2}}  \right),& \\
    &c_{1}r_{0}^{3}\omega_{0}^{3}+\frac{c_{2}}{r_{0}\omega_{0}} + \frac{c_{2}r_{0}\omega_{0}^{3}}{g^2}\leq P_{\text{lim}}.&
\eea

We can simply find $r_{0}$ which maximizes the minimum rate in (P1) and (P3) under constraints (40) and (41) via one-dimensional line search. For the EE maximization problems (P2) and (P4), $r_{0}$ can be computed in the range of (40). As a result, the initial trajectory $\qv_{0}[n]$ can be written by $\qv_{0}[n] = [r_{0}\cos2\pi\frac{n}{N} + x_{0} \ \ r_{0}\sin2\pi\frac{n}{N} + y_{0} ]^{T} \ (n=0,1,...,N)$ and the initial velocity $\vv_{0}[n]$ can be simply obtained as $\vv_{0}[n]=(\qv_{0}[n+1]-\qv_{0}[n])/\delta_{t} \ (n=0,1,...,N-1)$ assuming $\delta_{t}^{2}\approx0$ in (2).

\section{Numerical Results} \label{sec:numerical results}
In this section, we provide numerical results to validate the effectiveness of the proposed algorithms. For the simulations, we consider $K=6$ GNs which are distributed as in Fig. 2 where the locations of the GNs are marked with the triangles. The constant altitude, the bandwidth, the reference SNR, and the peak transmission power are set to be $H = 100$ m, $W$ = 1 MHz, $\gamma_{0}$ = 80 dB, and $P_{\text{peak}} = $ 10 dBm, respectively.
Also, the minimum velocity, the maximum velocity, and the maximum acceleration of the UAV are determined as $V_{\text{min}} = 3$ m/sec, $V_{\text{max}} = 100$ m/sec, and $a_{\text{max}} = 5$ m/sec$^2$, respectively. For the propulsion power consumption model in (6), the constants $c_{1}$ and $c_{2}$ are set as $c_{1}$ =  9.26$\ \times \ $10$^{-4}$ and $c_{2}$ = 2250, respectively, which make the minimum propulsion power consumption $P_{\text{prop,min}}$ = 100 W when $\|\vv\|$ = 30 m/sec.

\begin{figure}
\begin{center}
\includegraphics[width=5in]{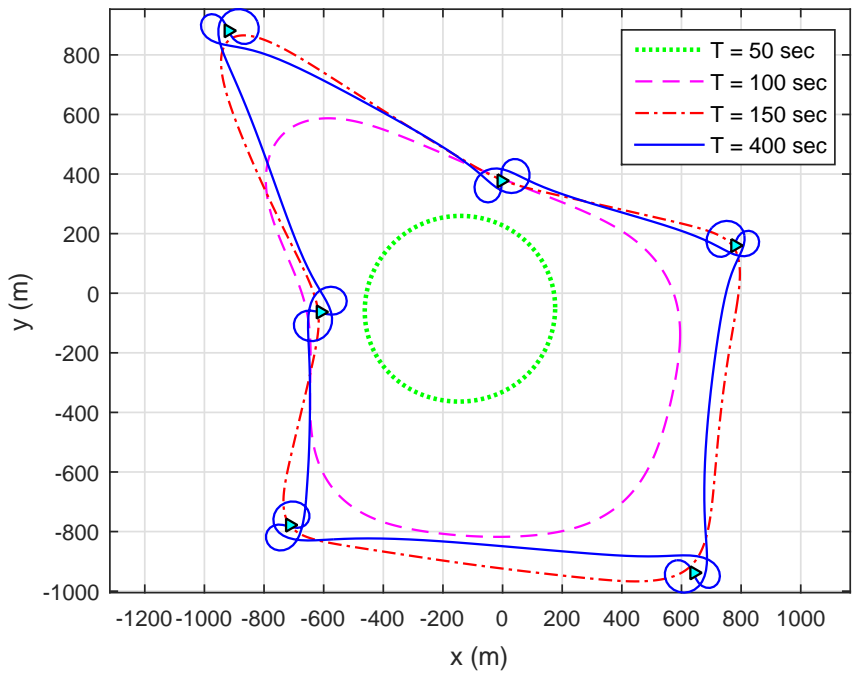}
\end{center}
\vspace{-3mm}
\caption{Optimized UAV trajectories for different periods $T$ with $P_{\text{lim}}=150$ W.}
\end{figure}
We first demonstrate the performance of the minimum rate maximization algorithms. Fig. 2 illustrates the optimized UAV trajectories with various $T$ for $P_{\text{lim}}$ = 150 W. It is observed that when $T$ is smaller than 150 sec, as $T$ increases, the UAV tries to get closer to all GNs in order to improve the channel conditions from the GNs. In contrast, if $T$ is sufficiently large ($T$ = 400 sec), the UAV is now able to visit all the GNs within a given time period. Thus, the UAV can hover over each GN for a while by traveling smooth path around the GNs.
This is different from the results in \cite{QWu:17} where the UAV does not have practical movement constraints. This can be explained as follows: Due to the constraints on the velocity and the propulsion power, the UAV cannot stay at fixed positions as in \cite{QWu:17}. Therefore, the UAV continuously moves around as close to the GNs as possible to maintain good communication channels without exceeding the propulsion power limit $P_{\text{lim}}$.

\begin{figure}
\begin{center}
\includegraphics[width=5in]{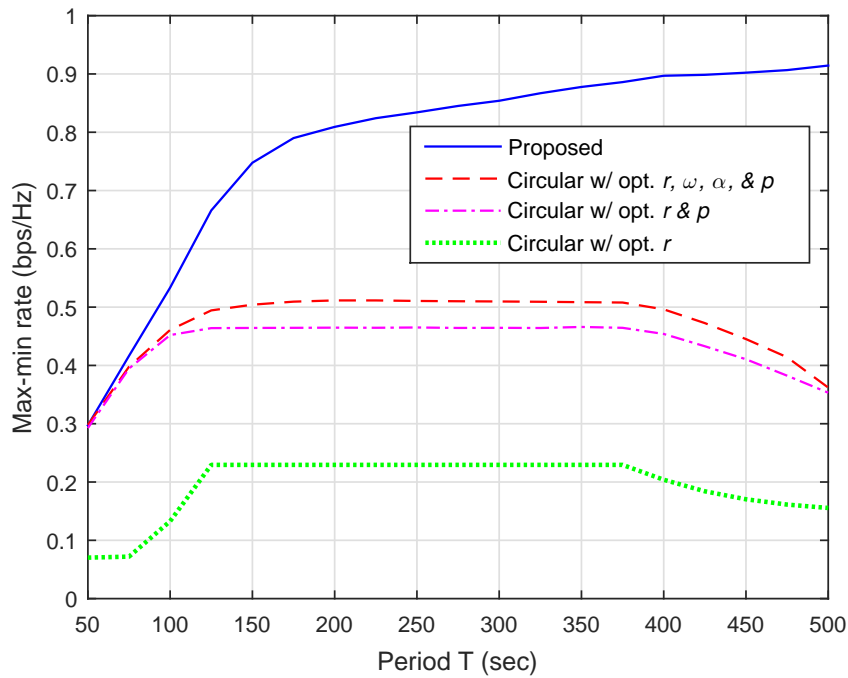}
\end{center}
\vspace{-3mm}
\caption{Max-min rate with respect to the period $T$ with $P_{\text{lim}}=150$ W.}
\end{figure}

Fig. 3 shows the maximized minimum (max-min) rate performance of the proposed algorithm as a function of $T$. We compare the performance of the proposed algorithm with the following circular trajectory based methods.
\begin{itemize}
\item[-] \textit{Circular with optimum $r, \ \omega, \ \alpha,$  and $p$}: radius, angular velocity, angular acceleration, and uplink transmit power are jointly optimized with (P3) in Section \rom{4}-A with the circular trajectory.
\item[-] \textit{Circular with optimum $r$ and $p$}: radius and uplink transmit power are jointly optimized with (P3.1) in Section \rom{4}-A with the circular trajectory.
\item[-] \textit{Circular with optimum $r$}: radius is optimized with $P_{\text{peak}}$ as the initial circular trajectory in Section \rom{4}-B
\end{itemize}
First, it can be verified that the proposed algorithm outperforms the baseline schemes regardless of the time period $T$. Also, we can see that the max-min rate in the proposed algorithm monotonically increases with $T$, since more time is available at the UAV to hover around each GN. In contrast, in the baseline schemes which are restricted in circular shape trajectory, the max-min rate performance first increases as $T$ grows, and then decreases after a certain $T$. This is due to a fact that in order to satisfy the propulsion power constraint, the radius of the circular trajectory should increase as $T$ gets large, and thus the UAV may become too far away from the geometric center of the GNs after a certain $T$. Therefore, we can expect the performance gain of the proposed algorithm over baseline schemes is to grow with $T$.

\begin{figure}
\begin{center}
\includegraphics[width=5in]{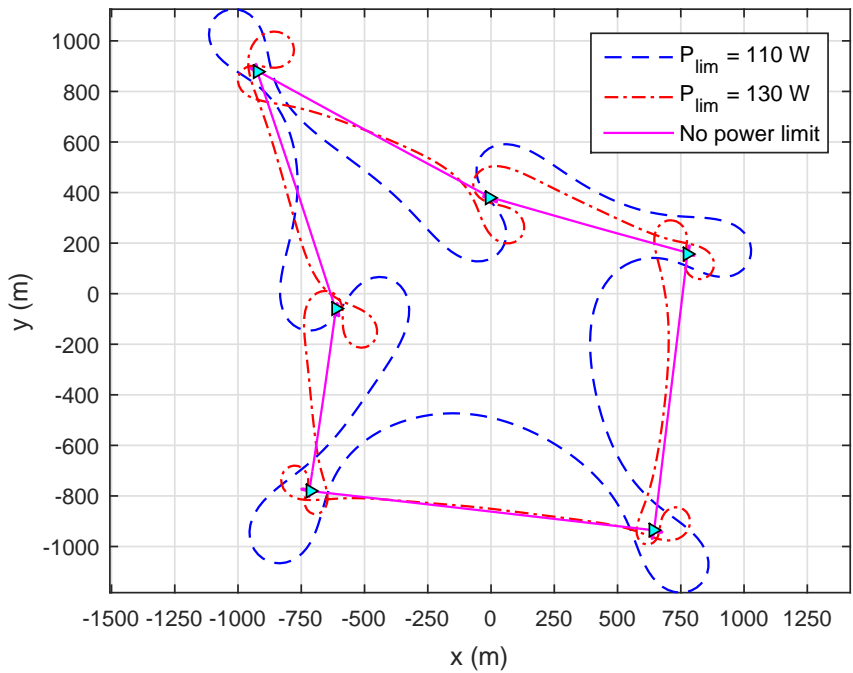}
\end{center}
\vspace{-3mm}
\caption{Optimized UAV trajectories for different propulsion power limit $P_{\text{lim}}$ with $T=400$ sec.}
\end{figure}

Fig. 4 illustrates the optimized UAV trajectories for various propulsion power limit $P_{\text{lim}}$ with $T$ = 400 sec. It can be shown that for $P_{\text{lim}}$ = 110 W, the trajectory of the UAV is restricted to a smooth path with a large turning radius to consume a low propulsion power. However, as $P_{\text{lim}}$ gets larger, we observe quick changes along the trajectory path.
Thus the UAV can move with a much smaller turning radius, which enhances the max-min rate performance.

\begin{figure}
\begin{center}
\includegraphics[width=5in]{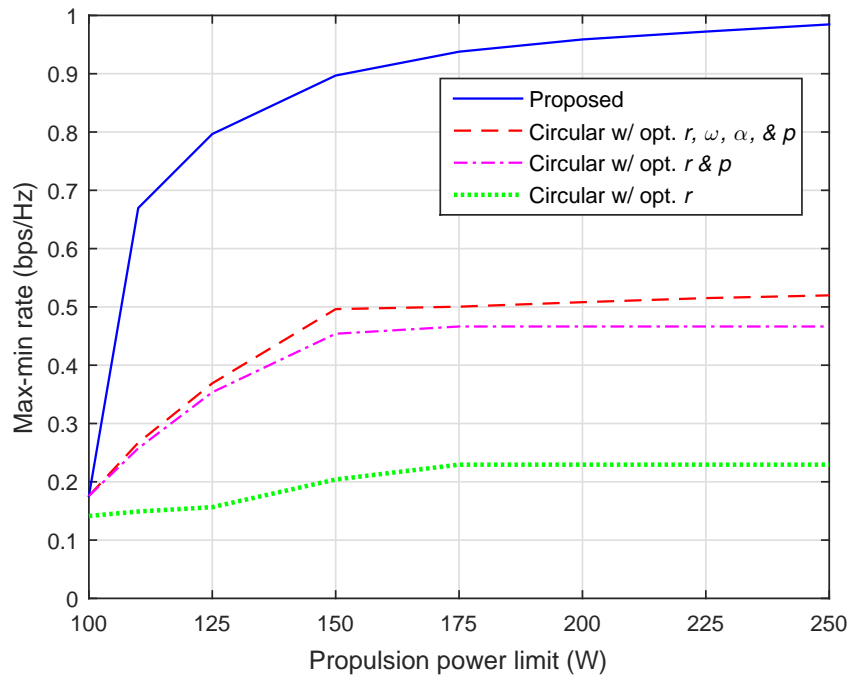}
\end{center}
\vspace{-3mm}
\caption{Max-min rate with respect to the propulsion power limit $P_{\text{lim}}$ with $T=400$ sec.}
\end{figure}

In Fig. 5, we depict the average max-min rate of various schemes as a function of the propulsion power constraint $P_{\text{lim}}$. For both the proposed algorithm and the baseline schemes, the max-min rate first increases as $P_{\text{lim}}$ grows and then gets saturated. This can be explained as follows: With a large $P_{\text{lim}}$, the trajectory and the velocity of the UAV change more freely to attain good channel conditions, and thus the max-min rate increases. However, even if a large $P_{\text{lim}}$ is given, the max-min rate cannot continue to increase because there are practical limits on the velocity and acceleration. Similar to Fig. 3, we can see that the proposed algorithm provides significant performance gains over the baseline schemes.

\begin{figure}
\begin{center}
\includegraphics[width=5in]{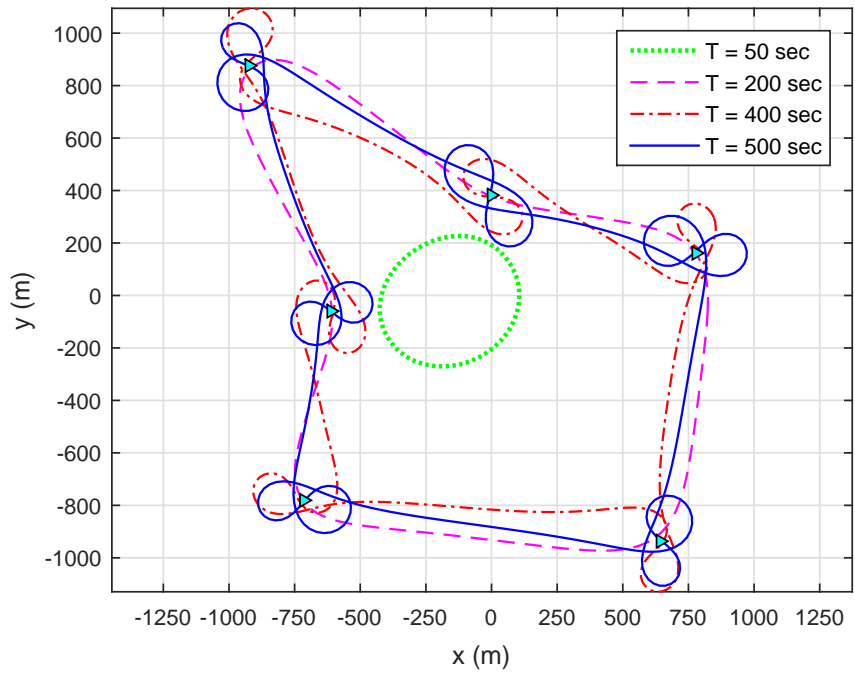}
\end{center}
\vspace{-3mm}
\caption{Optimized energy efficient UAV trajectories for different periods $T$}
\end{figure}
Next, in Fig. 6, we investigate the optimized trajectory of the EE maximization problem with various $T$. As $T$ increases, the overall patterns are similar to Fig. 2. Nevertheless, to balance between the rate performance and the propulsion power consumption, the EE maximization trajectory shows a smooth path with a relatively large turning radius, and thus the average propulsion power consumption becomes lower.

%

\begin{figure}
\begin{center}
\includegraphics[width=5in]{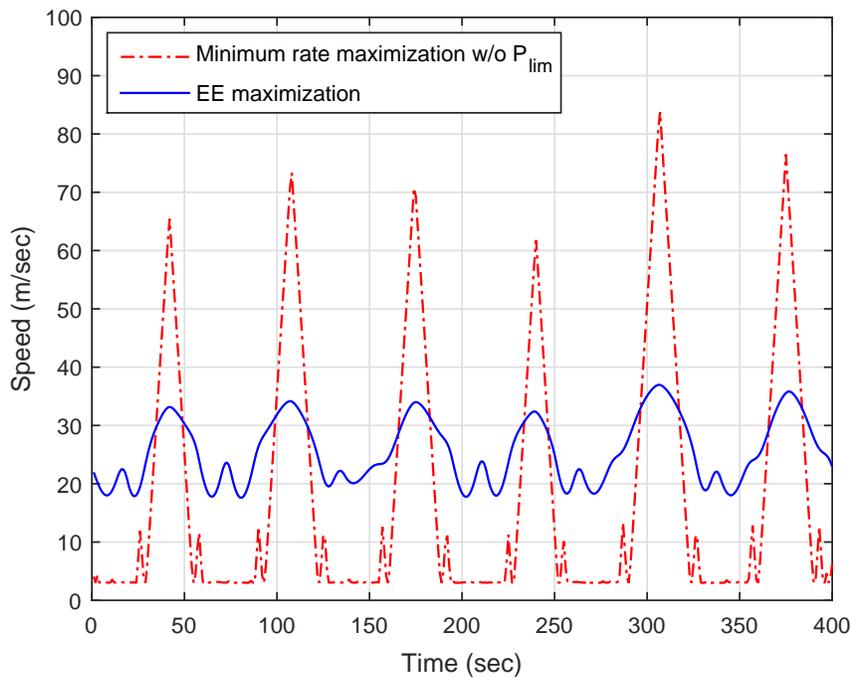}
\end{center}
\vspace{-3mm}
\caption{UAV speeds for the max-min rate without propulsion power constraint and the EE maximization with $T$ = 400 sec}
\end{figure}

To present the impact of the energy efficient UAV communication designs, Fig. 7 depicts the UAV speed of the proposed EE maximization method with $T$ = 400 sec. For comparison, we also consider the max-min rate scheme without the propulsion power constraint.
It is observed that for the max-min rate case, the UAV tries to fly between the GNs as fast as possible and stay over the GNs with a low speed. On the other hand, the EE maximization scheme keeps the speed of the UAV at around 30 m/sec in order not to waste the propulsion energy.

\begin{table}
\centering
\caption{Performance comparison with max-min rate and EE maximization for $T$ = 400 sec}
\begin{tabular}[ht!]{|c|c||c|c|c|c|c|}
\hline
\multicolumn{2}{|c||}{} & Average   & Average   & Average  & Average  & Energy   \\
\multicolumn{2}{|c||}{} &speed   &acceleration & max-min rate &power & efficiency  \\
\multicolumn{2}{|c||}{} &     (m/sec)        & (m/sec$^{2}$)&(bps/Hz) &(Watts) &(kbits/Joule) \\ \hhline{|=|=||=|=|=|=|=|} 
Max-min rate &  Proposed &18.42  & 4.73& 0.99  &  553.01& 1.80 \\ \cline{2-7}
 w/o $P_{\text{lim}}$ & Circular &13.50  & 1.88& 0.53  &  541.41&0.98 \\ \cline{1-7}
EE maximization & Proposed & 25.73  & 2.71&0.79 &122.14 &6.47 \\ \cline{2-7}
\ & Circular & 15.35  & 0.27&0.47 &151.33 &3.10 \\
\hline
\end{tabular}
\end{table}
Finally, Table \rom{1} presents the performance comparison of the max-min rate without propulsion power constraint and the EE maximization designs for both the proposed and the baseline schemes with $T$ = 400 sec. We can see that the max-min rate methods consume much higher propulsion power by allowing a large variation of the speed and the average acceleration. In contrast, the speed of the proposed EE maximization design slowly varies with low acceleration, and thus much higher EE can be achieved. We observe that the proposed EE maximization algorithm exhibits about 259 $\%$ gain over the max-min rate without the propulsion power constraint and 109 $\%$ gain over the circular baseline EE maximization scheme.

\section{Conclusion} \label{sec:conclusion}
In this paper, we have studied the UAV-aided wireless communication optimization under the practical propulsion energy constraint at the UAV. For both the minimum average rate maximization problem and the EE maximization problem, the UAV trajectories and the uplink transmit power of the GNs have been jointly optimized. By applying the SCA technique, we have proposed efficient iterative algorithms which find local optimal solutions. Numerical results have demonstrated that the proposed algorithms provide substantial performance gains compared to the baseline schemes.

\begin{appendices}

\section{proof of lemma 1}
Let us define a function $f_{1}(\uv) \triangleq \frac{1}{ {\rho\| \uv \| }^{2} + z}$ for $\uv=[u_{x} \ u_{y}]^{T} $ where $z$ and $\rho$ are positive constants. For any given $\uv_{l}  \in\mathbb{R}^{2\times1} $, in order for arbitrary function $g_{1}(  \uv | \uv_{l} )$ to be a concave surrogate function of $f_{1}(\uv)$, it must satisfy the following conditions: $f_{1}(\uv_{l}) =g_{1}(  \uv_{l} | \uv_{l} )$, $ \nabla g_{1}(\uv_{l}| \uv_{l}) = \nabla f_{1}(\uv_{l})$, and $g_{1}(\uv| \uv_{l}) \leq f_{1}(\uv), \forall \uv$ \cite{BMarks:78}. Denoting the function $g_{1}(  \uv | \uv_{l} )$ as
\bea
    g_{1}(  \uv | \uv_{l} ) \triangleq -\frac{\rho{\| \uv \| }^{2}}{z^2} + \overline{B} \uv^{T}\uv_{l} + \overline{C},
\eea
where $\overline{B} \triangleq 2\rho\left( \frac{1}{z^{2}} - \frac{1}{{\left(    {\rho\|\uv_{l} \|^{2} +z }     \right)    }^2}       \right)$ and $\ \overline{C} \triangleq \frac{1}{ \rho\|\uv_{l} \|^{2} +z} + \frac{2\rho\|\uv_{l} \|^{2}}{{\left(    {\rho\|\uv_{l} \|^{2} +z }     \right)    }^2}-\frac{\rho\|\uv_{l} \|^{2}}{z^2}$, it can be easily shown that $f_{1}(\uv_{l}) =g_{1}(  \uv_{l} | \uv_{l} )$, i.e., $g_{1}(  \uv | \uv_{l} )$ fulfills the first condition for the surrogate function.

Also, the gradient of $f_{1}(\uv)$ and $g_{1}(  \uv | \uv_{l} )$ with respect to $\uv$ can be respectively computed as
\bea
    \nabla f_{1}(\uv) = -\frac{-2\rho\uv}{{\left(    {\rho\|\uv \|^{2} +z }     \right)    }^2}, \\
    \nabla g_{1}(  \uv | \uv_{l} ) = -\frac{2\rho\uv}{z^2} + \overline{B}\uv_{l}.
\eea
Since two gradients in (43) and (44) become identical at $\uv = \uv_{l}$, $g_{1}(  \uv | \uv_{l} )$ satisfies the second condition for the surrogate function.

To prove the global lower bound condition, we can calculate the Hessian matrix $\nabla^{2}_{\uv}h_{1}(  \uv | \uv_{l} )$ of the function $h_{1}(  \uv | \uv_{l} ) \triangleq f_{1}(\uv) - g_{1}(  \uv | \uv_{l} )$ as
\bea
    \nabla^{2} h_{1}(  \uv | \uv_{l} ) = D \begin{bmatrix} E + 4\rho z^{2}u_{x}^{2} & 4\rho z^{2}u_{x}u_{y} \\ 4\rho z^{2}u_{x}u_{y} & E + 4\rho z^{2}u_{y}^{2} \end{bmatrix},
\eea
where $D \triangleq \frac{2\rho}{z^{2}{\left( \rho\|\uv \|^2 + z  \right)}^3}>0$ and $ E \triangleq \rho^{3}\|\uv \|^6 + 3\rho^{2}z\|\uv \|^{4} + 2\rho z^2\|\uv \|^{2} \geq 0. $
One can easily check that the Hessian in (45) is a positive semi-definite matrix, which implies that $h_{1}(  \uv | \uv_{l} )$ is a convex function.

Since $\nabla h_{1}(  \uv | \uv_{l} ) = \bf{0}$ at $\uv  = \uv_{l}$ from (43) and (44), the global minimum of $h_{1}(  \uv | \uv_{l} )$ is achieved at $\uv  = \uv_{l}$ with $h_{1}(  \uv_{l} | \uv_{l} )=0$.
As a result, we can show that $h_{1}(  \uv | \uv_{l} )$ is greater than or equal to 0 for any given $\uv_{l}$, and thus the third condition for the surrogate function holds. By substituting $\uv = \qv_{l+1}[n]-\wv_{k}$, $\uv_{l} = \qv_{l}[n]-\wv_{k}$, $z = H^{2}$, and $\rho=1$ and multiplying $f_{1}(\uv)$ and $g_{1}(\uv |\uv_{l})$ by $P_{\text{peak}}\gamma_{0}$, Lemma 1 is thus proved.

\end{appendices}

\bibliographystyle{ieeetr}
\bibliography{AZREF}

\end{document}